\RequirePackage{fix-cm}
\documentclass[twocolumn]{svjour3}          % twocolumn
\smartqed  % flush right qed marks, e.g. at end of proof
\usepackage{graphicx}
\usepackage{multirow}
\usepackage[caption=false]{subfig}
\usepackage{comment}
\usepackage{soul}
\usepackage{booktabs}
\usepackage{color}

\begin{document}
\sloppy
\title{Studying Leaders \& Their Concerns Using Online Social Media During The Times Of Crisis - A COVID Case Study%\thanks{Grants or other notes
}


\author{Rahul Goel         \and
        Rajesh Sharma
}


\institute{Rahul Goel \and Rajesh Sharma \at
              Institute of Computer Science, University of Tartu, Estonia \\
              \email{rahul.goel@ut.ee; rajesh.sharma@ut.ee}} 
%}

\date{Received: date / Accepted: date}

\maketitle

\begin{abstract}
Online social media (OSM) has emerged as a prominent platform for debate on a wide range of issues. Even celebrities and public figures often share their opinions on a variety of topics through OSM platforms. One such subject that has gained a lot of coverage on Twitter is the Novel Coronavirus, officially known as COVID-19, which has become a pandemic and has sparked a crisis in human history. In this study, we examine 29 million tweets over three months to study highly influential users, whom we refer to as leaders. We recognize these leaders through social network techniques and analyze their tweets using text analysis. Using a community detection algorithm, we categorize these leaders into four clusters: \textit{research}, \textit{news}, \textit{health}, and \textit{politics}, with each cluster containing Twitter handles (accounts) of individual users or organizations. E.g., the \textit{health} cluster includes the World Health Organization (@WHO), the Director-General of WHO (@DrTedros), and so on. The emotion analysis reveals that (i) all clusters show an equal amount of \textit{fear} in their tweets, (ii) \textit{research} and \textit{news} clusters display more \textit{sadness} than others, and (iii) \textit{health} and \textit{politics} clusters are attempting to win public \textit{trust}. According to the text analysis, the (i) \textit{research} cluster is more concerned with recognizing \textit{symptoms} and the development of \textit{vaccination}; (ii) \textit{news} and \textit{politics} clusters are mostly concerned with \textit{travel}. We then show that we can use our findings to classify tweets into clusters with a score of 96\% AUC ROC.

\end{abstract}

\section{Introduction}
A leader is a person who can reach and inspire a large number of people with her/his opinions \cite{goleman2004makes}. Crisis times provide an opportunity to test these leaders and assess their ability to take timely and necessary actions \cite{littlefield2007crisis,schroeder2012mining,bruns2013arab,littlefield2007crisis}. Nowadays, leaders often use online social media (OSM) platforms to communicate information to crisis-affected people efficiently. During a crisis, such as the current COVID-19 pandemic, communication is critical since it influences people's opinions, attitudes, and psychological conditions. Many public figures worldwide have acknowledged that the infodemic of disinformation is a significant secondary crisis caused by the pandemic, including UN Secretary-General \cite{pastor2020characterizing}. Infodemics can exacerbate the pandemic's real-world adverse effects across many dimensions, including social, physical, and even sanitary. Thus, leaders often ensure that accurate information is widely disseminated and that misinformation is clarified using OSM.

This work centered on Twitter as it has attracted more than 628 million tweets related to COVID-19 by May 8, 2020\footnote{https://www.tweetbinder.com/blog/covid-19-coronavirus-twitter/} primarily because the COVID-19 pandemic affected more than 180 countries and territories worldwide \cite{csse2020coronavirus}. It impacted the lives of many individuals by keeping them in isolation or total lockdown. During these conditions, individuals are inclined to use social media platforms to share their experiences or concerns.
\subsection{Study Highlights}

In this paper, we analyze 29 million tweets from 6 million unique users, collected through Twitter's API by using COVID-19 trending keywords on Twitter (``coronavirus", ``coronovirusoutbreak" and ``COVID-19"). Our dataset spans from February 01, 2020, to May 02, 2020, and this study covers the following:
\begin{enumerate}
    \item \textbf{Identifying leaders: }We recognize the leaders during the COVID-19 crisis by creating users' tweet network and using social network analysis techniques. We define \textit{Leaders} as users with high PageRank values. Please note that \textit{leaders} can either be an individual user or an organization (e.g., the \textit{World Health Organization}). However, we consider them as a separate entity and do not distinguish between them. We use Girvan–Newman community detection method to identify the optimal number of communities. Based on the highest modularity value, we categorize leaders into four clusters\footnote{Note: In this work, ``leader" represents a  Twitter handle, and ``cluster" represents a group of Twitter handles.}: \textit{research}, \textit{news}, \textit{health}, and \textit{politics} (refer Section \ref{Sec:Leaders} for detail). We further study these clusters by analyzing their tweets to reveal their inclination and concerns during COVID-19.
    \item \textbf{Leader's concerns: } Our finding shows that leaders are mostly discussing about five topics that are related to (1) \textit{symptoms}; (2) \textit{vaccination}; (3) \textit{hygiene}; (4) \textit{travel}; and (5) \textit{pandemic} (Section \ref{Sec:Leaders}). We also use various text analysis techniques to understand clusters' alignment towards these concerns during the COVID-19 crisis. We observe that \textit{research} cluster is more concerned about understanding the \textit{symptoms} and development of \textit{vaccination}; \textit{news} and \textit{politics} clusters about \textit{travel} and \textit{hygiene}. Finally, using Chi square independent test, we show that the number of concerns (or topics) and the type of leading cluster are dependent (Section \ref{Sec:leadersConcern}).
    \item \textbf{Classification model: } Based on our findings, we build a multi-class predictive model for estimating the probability that a tweet belongs to a specific leading cluster. Our model can classify tweets among leading clusters with a score of 96\% AUC ROC (Section \ref{Sec:Modeling}).
\end{enumerate}

\subsection{Contributions}
The contribution of our analysis is three-fold:
\begin{enumerate}
    \item Previous COVID-19 studies of leadership used social media to propose a framework to characterize leaders based on nodes centrality \cite{pastor2020characterizing,rufai2020world}. In contrast, this paper identifies leaders’ and their clustering using community detection algorithms and the PageRank centrality method.
    \item Next, we show that the number of public concerns (or topics) and the type of leading cluster is dependent on one another using Chi-square independent test with a p-value of 1.411e-33.
    \item Finally, while previous research has only identified leaders, we use a machine learning approach to classify tweets into relevant clusters using tweets' features such as tweet's text, its sentiment, and topic.
\end{enumerate}

\subsection{Paper Organization}
The rest of the paper is as follows. Next, we discuss the related work. We then describe the dataset in Section \ref{Sec:Dataset}. Section \ref{Sec:Leaders} presents the study of leaders during COVID-19 and their alignments towards various concerns are covered in Section \ref{Sec:leadersConcern}. In Section \ref{Sec:Modeling}, we use various machine learning models to classify tweets into various leading clusters with good accuracy. We conclude with a discussion of future directions in Section \ref{Sec:Conclusion}.

\section{Related Work}\label{sec:RelatedWork}
Our work lies at the intersection of leadership, social media, and the COVID-19 pandemic. Thus, in this section, we present literature concerning various works which have studied leadership from different viewpoints. Broadly, studies on leadership types during COVID-19 can be divided into the following: (1) Religious leaders, (2) Academic leaders, and (3) Information leaders.

\subsection{Religious Leaders During Covid-19}
Religious leaders of faith-based communities played a crucial role in supporting individuals, families, and communities in coping with the pandemic, especially those who were ill or bereaved by COVID-19 \cite{greene2020psychological}. To acknowledge that COVID-19 is a global pandemic, affecting all races, ethnicities, and geographic regions, various global religious and inter-religious groups have issued guidance, advisory notes, and statements to support the actions and roles of religious leaders during the pandemic \cite{world2020practical}. In \cite{yezli2021covid}, the authors believe that temporary closure of places of worship for group prayers and religious services should be implemented worldwide (especially in countries with local COVID-19 transmission) regardless of faiths involved.

In another work, the authors describe the religiously innovative adaptations made to customary rituals by Jewish religious leaders to fight COVID-19 \cite{frei2020going}. These adaptations included allowing spiritual prayer through a "balcony", conducting online prayer sessions using video conferencing, and broadcasting the Passover ceremony. In \cite{dein2020covid}, the authors examined anxiety and distress among members of a Modern Orthodox Jewish community to be quarantined in the USA. Their study demonstrates that public health authorities have an opportunity to form partnerships with religious leaders to promote health and psycho-social support, protect communities against stigma and discrimination.

Additionally, several works discussed the impact of virtual religious activities on mental health \cite{dein2020covid}. The authors showed that limited empirical research is available on religion and mental health, and they have relied heavily on newspaper and internet sources. It has also been studied that healthcare professionals are often unprepared to answer the patients' religious beliefs regarding the diseases \cite{hashmi2020religious}, which results in religious clichés and stigma among patients. Therefore, the inclusion and collaboration of spiritual leaders with healthcare professionals are needed to ensure a holistic understanding and overcome the stigma that can shape as a barrier for reaching an optimal therapeutic outcome \cite{hashmi2020religious}.

\subsection{Academic Leaders During Covid-19}
Academic leaders worldwide have responded by moving their educational and associated activities online; as a sense of immediacy swept the nation. The decision to pivot to remote learning was made swiftly, particularly by those institutions operating a shared leadership model \cite{fernandez2020academic}. A survey study sought to understand the impact of the COVID-19 pandemic on school leadership in K-12 schools in a South Texas region indicate that school leaders were generally confident in their preparedness to best serve students, staff, and parents during the COVID-19 pandemic but felt a lack of resources and a preponderance of student inequities complicated the experience \cite{varela2020leading}.

In \cite{brammer2020covid}, the authors discussed that COVID‐19 has profound impacts on tertiary education globally. Border closures, cuts to aviation capacity, mandatory quarantine on entering a country, restrictions on mass gatherings, and social distancing all pose challenges to higher education institutions. However, there are possible solutions for Academic Leadership to improve diversity, equity, and inclusion (DEI) in the workplace, at community and institute levels, and in broader policy and decision-making \cite{maas2020academic}.

\subsection{Information Leaders During Pandemics}
In the past, researchers have analysed Twitter data for studying various epidemic outbreaks such as Ebola \cite{oyeyemi2014ebola,carter2014twitter}, H1N1 \cite{smith2010little}, flu \cite{achrekar2011predicting}, swine flu \cite{ritterman2009using} etc. In \cite{chew2010pandemics}, the authors illustrate the potential of monitoring public health using social media during the H1N1 pandemic. In another work \cite{signorini2011use}, authors track the disease level and public concern during the H1N1 pandemic in the US using Twitter data.

A set of work has also focused on understanding the epidemic spreading patterns at different geographic locations using social media data and call record data \cite{goel2020mobility,goel2019modeling}. They also studied effect of COVID on stock market\cite{goel2020covid} using social media data. In \cite{jain2015effective}, authors track the spread of influenza-A (H1N1) in India using Twitter data. Vaccination and anti-viral uptake during H1N1 in the UK are studied in \cite{mcneill2016twitter} using tweets. In \cite{kumar2018deadly}, the authors examined the Nipah virus in India. In a very recent work \cite{samaras2020comparing}, researchers compared Google and Twitter data for predicting influenza cases in Greece.

Information is vital during a crisis such as the current COVID-19 pandemic as it significantly shapes people's opinion, behavior, and even their psychological state. The Secretary-General of the United Nations has acknowledged that the infodemic of misinformation is an essential secondary crisis produced by the pandemic. Infodemics can amplify the pandemic's real negative consequences in different dimensions: social, economic, and even sanitary \cite{pastor2020characterizing}. For instance, infodemics can lead to hatred between population groups that fragment the society, influencing its response or result in negative habits that help the pandemic propagate. On the contrary, reliable and trustful information along with messages of hope and solidarity can be used to control the pandemic, build safety nets and help promote resilience and anti-fragility. Several works proposed frameworks to characterize Twitter leaders based on the social graph analysis derived from the activity in this social network \cite{pastor2020characterizing,rufai2020world,lee2020policy,goel2021studying}.

In this work, we analyzed online social media, particularly Twitter, to identify leaders using centrality and community detection algorithms during COVID-19. Additionally, we grouped these leaders into four major clusters (health organizations, politics, news, and research) and analyzed each cluster's tweet to identify their respective concerns using natural language processing technique, specifically using Latent Dirichlet Allocation (LDA) method with Gibbs sampling. Finally, we created a classification model using a machine learning technique to cluster tweets.

\section{Dataset Description}\label{Sec:Dataset}
This section first provides information about the coronavirus (officially known as COVID-19), and Twitter, an online social media platform. Next, the data collection and pre-processing are discussed in the following subsections.

\subsection{COVID-19 and Social Media}
In the last decade, humanity has faced many different pandemics such as  SARS, H1N1, and presently COVID-19 caused by the SARS-CoV-2 coronavirus~\cite{goel2020mobility}. The severity of these pandemics can be understood by the death toll claimed by them. The COVID-19 pandemic, which started in December 2019 from Wuhan \cite{world2020novel,huang2020clinical}, China has infected 128,253,282 individuals and claimed 2,804,677 (as of March $30^{th}$, 2021) deaths worldwide \cite{csse2020coronavirus,covid19global}. This impacted individuals life by keeping them in isolation, or total lockdown. During these lockdown conditions, individuals are inclined to use social media platforms to share their thoughts and experiences. By May 8, 2020, Twitter has attracted more than 628 million tweets related to coronavirus which makes it a significant online social media platform for COVID-19 study\footnote{https://www.tweetbinder.com/blog/covid-19-coronavirus-twitter/}.

\subsection{Twitter}
Twitter is an online social media platform classified as a microbloging site with which users can share messages, links to external Web sites, images, or videos that are visible to their followers. Messages that are posted are short in contrast to traditional blogs. Blogging becomes `micro' by shrinking it down to its bare essence and relaying the heart of the message and communicating the necessary as quickly as possible in real-time. Twitter, in 2016, limited its messages to 140 characters~\cite{giachanou2016like} and presently the maximum length of its message (also called a tweet) has been increased to 280 characters\footnote{https://developer.twitter.com/en/docs/basics/counting-characters}. Apart from Twitter, there are other microblogging platforms such as Tumblr\footnote{https://www.tumblr.com/}, FourSquare\footnote{https://foursquare.com/}, Google+\footnote{http://plus.google.com}, and LinkedIn\footnote{http://linkedin.com/} of which Twitter is the most popular microblogging site. In the past, researches have shown that Twitter data is well suited for sentiment analysis and opinion mining~\cite{tumasjan2010predicting}.

\subsection{Data Collection From Twitter}
\begin{figure}
    \centering
    \includegraphics[width=\columnwidth]{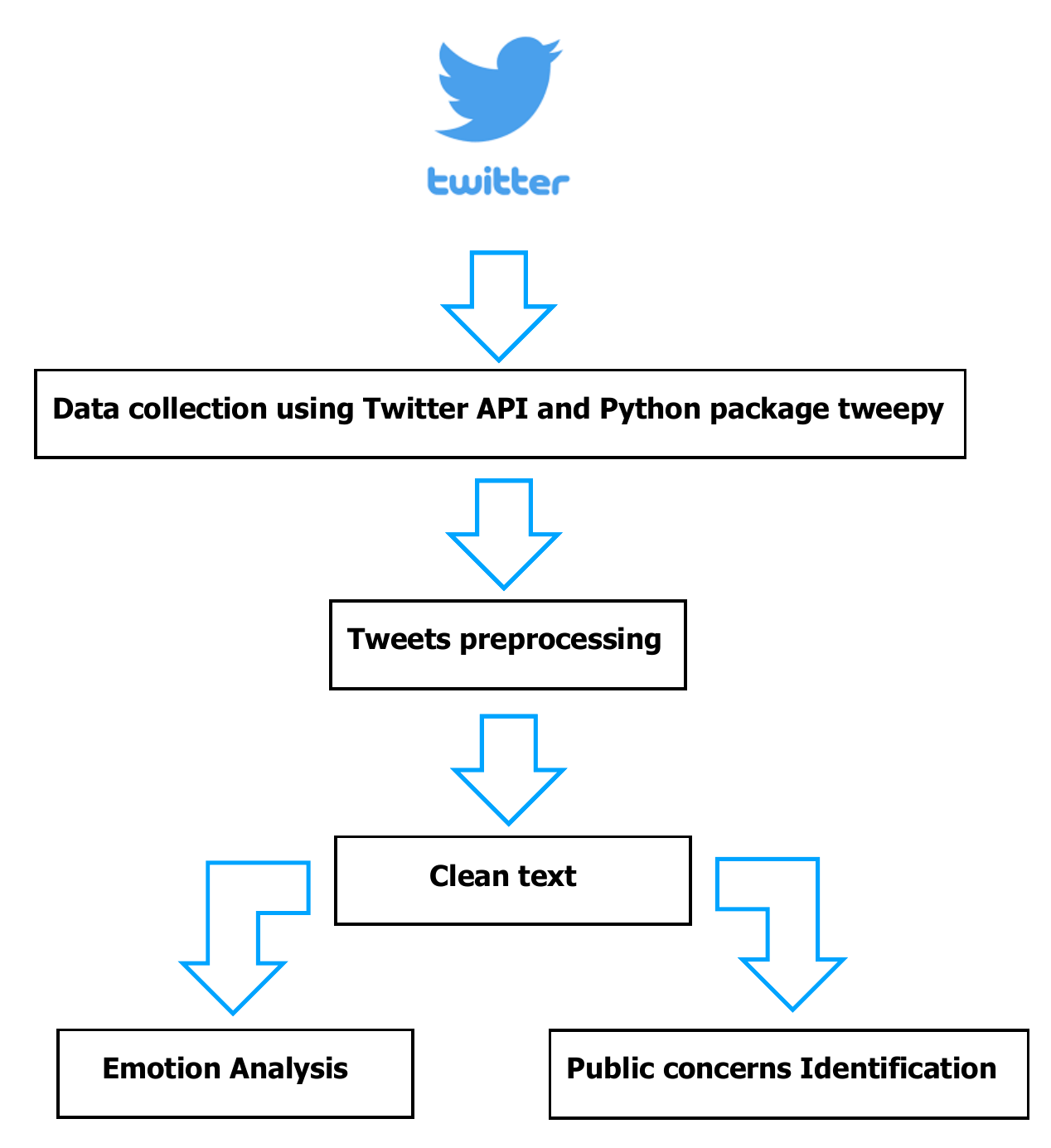}
    \caption{Tweets collection, preprocessing, emotions and public concern identification flow.}
    \label{fig:flow1}
\end{figure}

The data collection is done using Twitter Streaming Application Programming Interface (API) and Python (see Figure \ref{fig:flow1}). API is a tool that facilitates the interaction between computer programs and Web services. It enables the real-time collection of data by tracking the live stream of public tweets. Many Web services provide developers with APIs for interacting with their services, and for programmatically accessing data. Python library such as \textit{tweepy}\footnote{https://www.tweepy.org/} also assist this task by providing functions that can track live streams of public tweets using hashtags or usernames.

\begin{figure*}
    \centering
    \includegraphics[width=2\columnwidth]{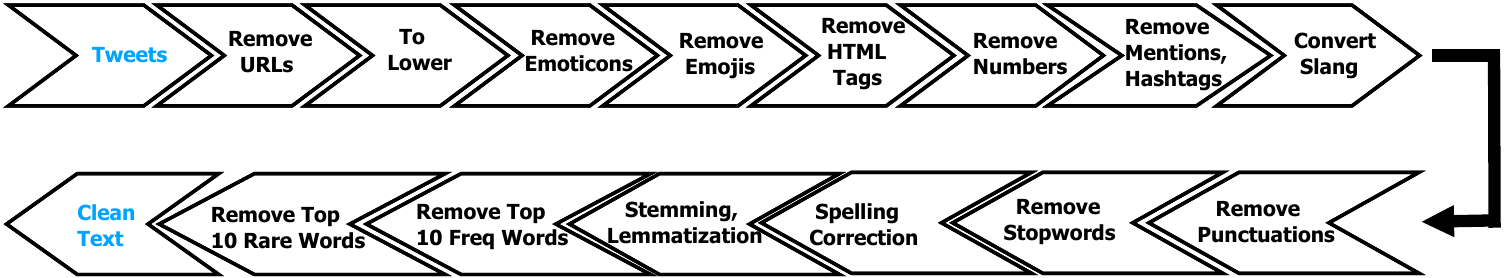}
    \caption{Tweets preprocessing.}
    \label{fig:flow2}
\end{figure*}

In this work, we use the Twitter Streaming API and Python package \textit{tweepy} to download tweets related to various keywords regarding rapidly increasing coronavirus disease (COVID-19). We collect and store a large sample of public tweets beginning February 1, 2020, that matched a set of pre-defined keywords: ``coronavirus" and ``coronovirusoutbreak". The additional keyword is later added such as ``COVID-19" (On February 11, 2020, COVID-19 was the official name given to novel coronavirus by WHO). The objective of using these specific keywords is to collect tweets belonging to COVID-19. The Twitter data is stored in JSON format to make it easy for parsing and analysis.

\subsection{Tweets Preprocessing}\label{subsec:DataPreprocessing}
Figure \ref{fig:flow2} shows the basic steps we took in preprocessing the tweet dataset. We start our tweet preprocessing by removing all the URL's as they do not include any useful information for text analysis. Afterward, the tweets are subjected to lowercasing. The emojis, emoticon, numbers, mentions (@) and hashtags (\#) are excluded from the tweets due to their specific semantics in tweets and overall text. In addition, all slang text are converted into its real meaning to understand the real context of the tweet.

Furthermore, we proceeded with removing every punctuation marks and a variety of different stopwords available in \textit{nltk} package. We also perform the spell correction using \textit{SpellChecker} package in Python. Using \textit{nltk} package, we did the stemming and lemmatization of the tweet text. The dataset's size in terms of total number of tweets, original tweets, retweets, number of distinct users, and the number of distinct features are shown in Table \ref{table:datasetStats}. Twitter features include information about tweets and users. This encapsulates core attributes that describe each tweet data such as author, message, unique ID, a timestamp of when it was posted, and sometimes geo metadata shared by the user. Each user information like Twitter name, ID, number of followers, and most often account biography are also associated with tweets. With each tweet we also get the tweet's common contents such as hashtags, mentions, media, and links.

\begin{table}
\begin{center}
  \begin{tabular}{ | l | c |}
    \hline
    \textbf{Parameter} & \textbf{Value}  \\ \hline
    Time period & 01-02-2020 to 02-05-2020 \\ \hline
    \#Tweets & 29,469,349 \\ \hline
    \#Original tweets & 6,494,657 \\ \hline
    \#Retweets &  22,974,692 \\ \hline
    \#Users &  7,875,334\\ \hline
    \#Features & 91 \\ \hline
  \end{tabular}
  \caption{Description of the dataset.}
  \label{table:datasetStats}
\end{center}
\end{table}

\section{Studying Leaders During COVID-19 Crisis}\label{Sec:Leaders}
To understand the users' interactions and to identify leaders, we first build a network by capturing the user's social connections. We build the directed ``retweet" network among users where an edge \textit{a} $\longrightarrow$ \textit{b} indicates that user \textit{a} retweets user \textit{b}. To identify users with similar alignment, we grouped them into communities using the Girvan–Newman method. Furthermore, to identify important nodes in the network, we employ the PageRank algorithm \cite{page1999pagerank}, a well-known algorithm to characterize the centrality of nodes. PageRank reflects the importance of a node in the retweet network, and a higher PageRank value represents influential users who can spread their tweet content to a community much faster compared to users with lower PageRank value.

\begin{figure}[ht!]
\includegraphics[width=\columnwidth]{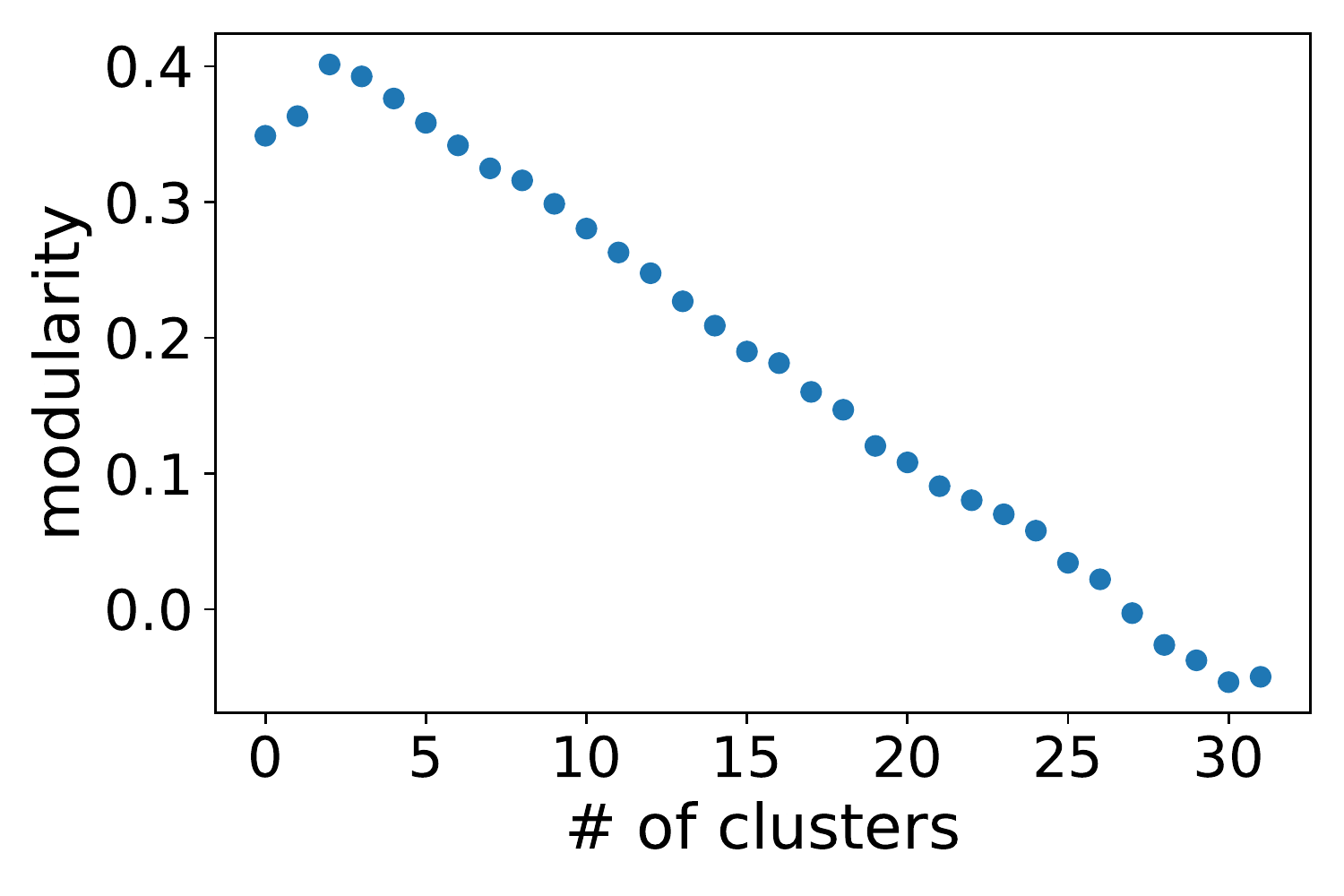}
\caption{\textbf{Optimal number of clusters using Girvan–Newman method.} Here, x-axis represents the number of clusters identified by Girvan–Newman community detection algorithm and y-axis represents modularity value for different numbers of clusters. We find that the optimal value for the number of clusters is 4 (with highest modularity value) for our dataset.}
\label{Fig:OptComm}
\end{figure}

The number of optimal communities identified by Girvan–Newman method with highest modularity is four (see Figure \ref{Fig:OptComm}). With some manual inspection of influential users in each community, we categorized communities into four clusters, \textbf{Health}, \textbf{Politics}, \textbf{News} and \textbf{Research}, depending upon their publicly available information such as profile description. Some of the examples of users under four different clusters are as follows:
\begin{enumerate}
    \item \textit{Health cluster}: which includes health related Twitter handlers such as \textit{World Health Organization (@WHO), Centers for Disease Control and Prevention (@CDCgov), Director General WHO (@DrTedros), Global Health Strategies (@GHS) and World Health Organization Western Pacific (@WHOWPRO)}.
    \item \textit{Politics}: includes Twitter profiles related to politicians such as \textit{U.S. President (@realDonaldTrump) and U.S. House Candidate}.
    \item \textit{News}: examples of Twitter handlers include \textit{China Global Television Network (@CGTNOfficial), Al Jazeera English (@AJEnglish)} and \textit{Global Times (@globaltimesnews)}.
    \item \textit{Research}: such as \textit{ISCR}.
\end{enumerate}

This indicate that, on Twitter, users are (re)sharing information from more reliable sources regarding COVID-19. Please note that some ``leaders'' are in fact teams, who tweet under a particular twitter handle, representing the official position of an office holder, e.g., Director General WHO(@DrTedros).

\noindent \textbf{Dominant emotions for various clusters: }
Next, we analyze tweets to discover each clusters' dominant emotions. As can be imagined, the majority of tweets from the Health, Politics, News and Research clusters would be fact-based, our dataset includes individuals' Twitter handles as well and these handles often express emotions that can be useful for our classification model. We use the \textit{Syuzhet}\footnote{https://github.com/mjockers/syuzhet} package~\cite{jockers2015syuzhet} available in R for emotions extractions. The \textit{Syuzhet} package is designed to work on a sentence level and repeated words do not count towards the emotion assignment. We manually inspected a random sample of tweets for the accuracy of \textit{Syuzhet} package and found it reasonably accurate with 83\% accuracy. We used original tweets (without pre-processing) for the manual checking.

\begin{figure}[ht!]
\includegraphics[width=\columnwidth]{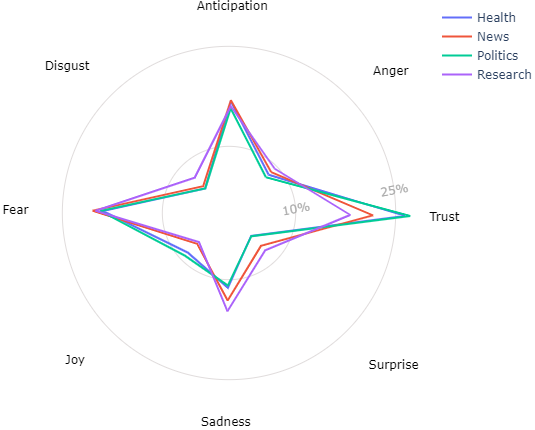}
\caption{Dominant emotions in tweets from various leaders.}
\label{Fig:Profession2}
\end{figure}

We observe that \textit{Anticipation}, \textit{Fear}, \textit{Sadness} and \textit{Trust} are dominant emotions for all clusters compared to \textit{Anger}, \textit{Disgust}, \textit{Joy} and \textit{Surprise} (see Figure \ref{Fig:Profession2}). Furthermore, it can be observe that all clusters show similar amount of \textit{fear} and \textit{anticipation} in their tweets. \textit{Political} and \textit{health} clusters indicate higher \textit{trust} compared to others. On the other hand, \textit{researchers} and \textit{news} clusters display greater \textit{sadness} towards the COVID-19 situation. Therefore, we can infer that leaders are displaying \textit{fear, anticipation} and \textit{sadness} but at the same time, they are trying to build a \textit{trust} environment among public.

\begin{figure*}[t]
\subfloat[February]{\includegraphics[width=0.3\linewidth]{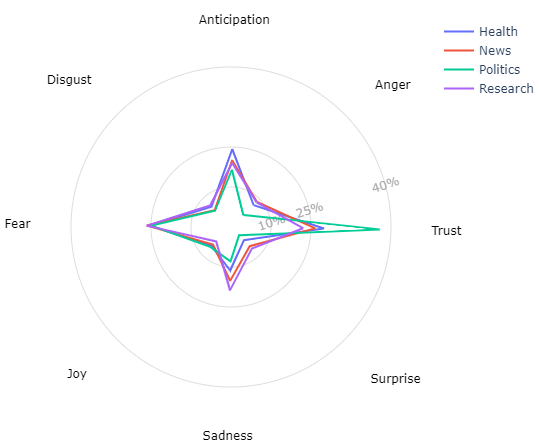}\label{fig:Emotions_months1}}
\hspace{3mm}
\subfloat[March]{\includegraphics[width=0.3\linewidth]{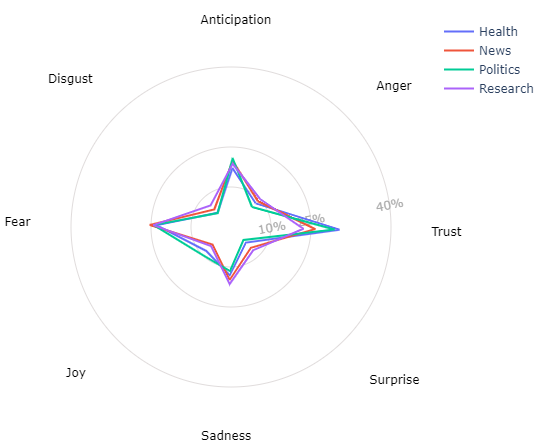}\label{fig:Emotions_months2}}
\hspace{3mm}
\subfloat[April]{\includegraphics[width=0.3\linewidth]{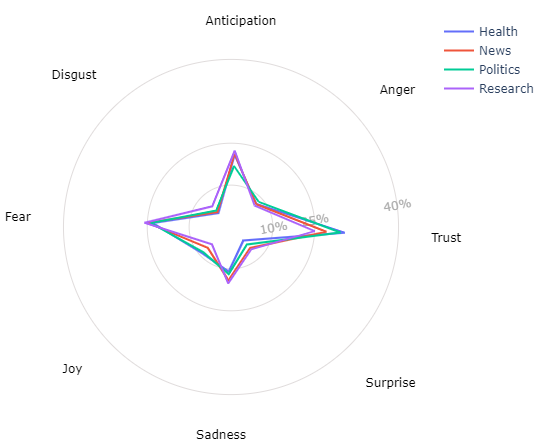}\label{fig:Emotions_months3}}
\caption{Dominant emotions in tweets from various leaders over time.}
\label{Fig:Emotions_months}
\end{figure*}

Furthermore, we analyzed the leaders’ dominant emotions over time. This is shown using Figure \ref{fig:Emotions_months1} (for February), \ref{fig:Emotions_months2}  (for March) and \ref{fig:Emotions_months3} (for April). We observe that there is only minor changes in emotions of all clusters, except in the politics cluster. In February, the politics cluster were showing more \textit{Trust}, but over time, it changed into \textit{Sadness} and \textit{Anticipation}. For other clusters, insignificant change in emotions is observed over time.

\noindent \textbf{Topic modelling: }In addition, we perform the text analysis to understand the focus of the interest of leaders during the COVID-19. We use the topic modeling technique to extract these interests. In particular, we use the Latent Dirichlet Allocation (LDA)~\cite{blei2003latent} method with Gibbs sampling~\cite{porteous2008fast} which is an unsupervised and probabilistic machine-learning topic modeling method that extracts topics from text data. The key assumption behind LDA is that each given text is a mix of multiple topics.  

We extracted between three to ten topics from tweets’ text and found that our dataset covers five different topics. To check this result and be sure on the number of topics, we compute two typical measurements used in LDA: perplexity and coherence scores for robustness check. The LDA returned a set of ten words related to each identified topic but not the title of the topic (see Table \ref{table:LDATopicOverall}, Column 2). %\fixme{2 brackets side by side. Can we do something about it?}. 
We then assign appropriate topics to each set of words that closely reflect the topic at an abstract level (Table \ref{table:LDATopicOverall}, Column 3). It can been observed that leaders are mostly discussing topics related to various public concerns during the COVID-19. These concerns are: (1) \textit{disease symptoms}, (2) \textit{disease vaccination}, (3) \textit{disease countermeasures or hygiene}, (4) \textit{disease transmission during travel} and (5) COVID-19 as \textit{pandemic/epidemic}. In the next section, we analyze the clusters' alignment toward these topics, which also reflects pubic concerns.

\begin{table}[t]
\begin{small}
\begin{center}
  \begin{tabular}{| p{0.07\columnwidth} | p{0.50\columnwidth} | p{0.27\columnwidth} |}
    \hline
    \textbf{S.No.} & \textbf{Words} & \textbf{Topic (Proposed)} \\ \hline
    1 & flu, symptomatic, fever, infected, spreading, china, cold, sick, severe, asymptomatic & Symptoms \\ \hline
    2 & china, effective, flu, cdc, world, research, testing, vaccine, pandemic, treatment & Vaccination\\ \hline
    3 & hand, wash, sanitizer, time, home, touch, masks, water, soap & Countermeasures \\ \hline
    4 & italy, international, quarantine, pandemic, traveling, restrictions, china, government, iran, world  & Travel \\ \hline
    5 & china, stop, cdc, who, pandemic, public, death, rate, flu, news & Pandemic \\ \hline
  \end{tabular}
  \caption{Topic Modelling using LDA with Gibbs sampling.}
  \label{table:LDATopicOverall}
\end{center}
\end{small}
\end{table}

\section{Clusters Alignment Towards Various Concerns}\label{Sec:leadersConcern}
To study the alignment of various clusters towards five most discussed topics or concerns discovered in Section \ref{Sec:Leaders} regarding rapidly increasing coronavirus disease (COVID-19), we annotate each tweet using specific keywords. For example, disease symptoms tweets are annotated (i.e., using keyword \textit{symptom}); vaccination (i.e., using keywords \textit{vaccine} and \textit{vaccination}); disease countermeasures (i.e., keywords \textit{hygiene}, \textit{wash}, \textit{hand} and \textit{mask}); disease transmission during travel (i.e., keywords \textit{travel}, \textit{flying}, \textit{fly}, \textit{airplane}, \textit{flight} and \textit{trip}) and pandemic (i.e., keywords \textit{pandemic} and \textit{epidemic}). In this section, we start by uncovering the specifics of these public concerns to understand them in more detail. We also explain the alignment of leading clusters towards these concerns.

\subsection{Disease symptoms}
\begin{figure*}[t]
\subfloat[Symptoms]{\includegraphics[width=0.23\linewidth]{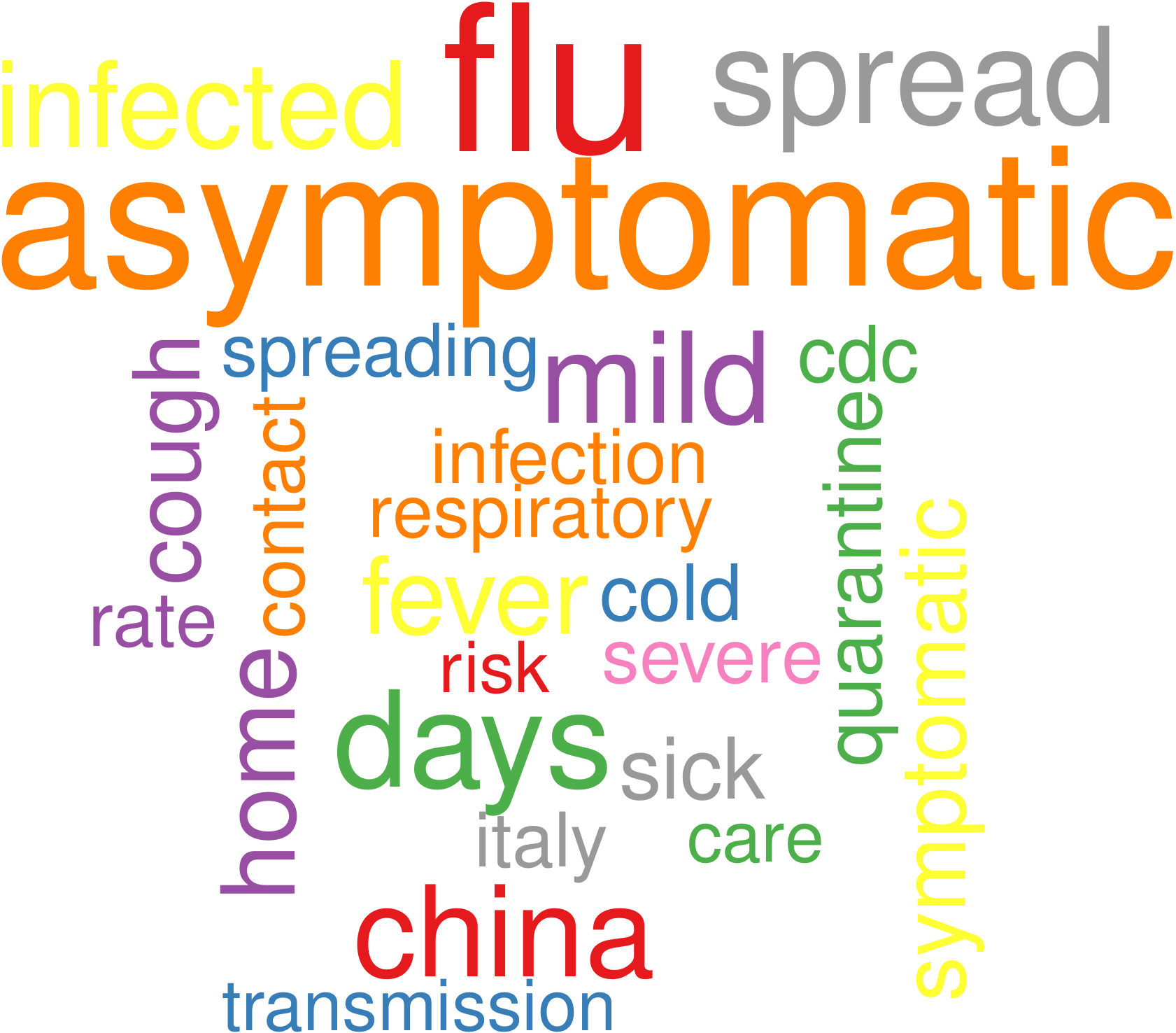}\label{fig:a}}
\hspace{3mm}
\subfloat[Vaccine]{\includegraphics[width=0.23\linewidth]{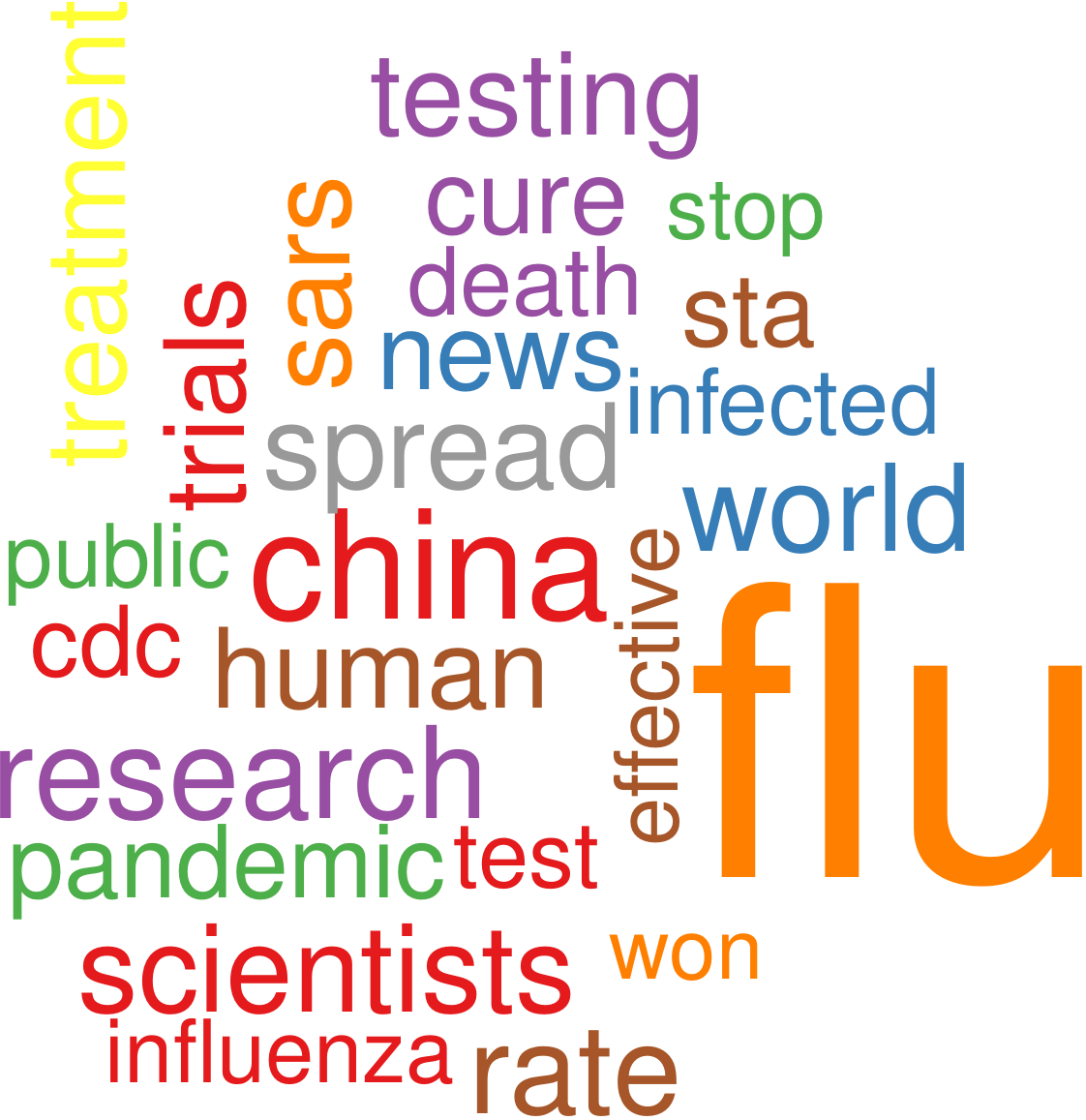}\label{fig:b}}
\hspace{3mm}
\subfloat[Countermeasures]{\includegraphics[width=0.23\linewidth]{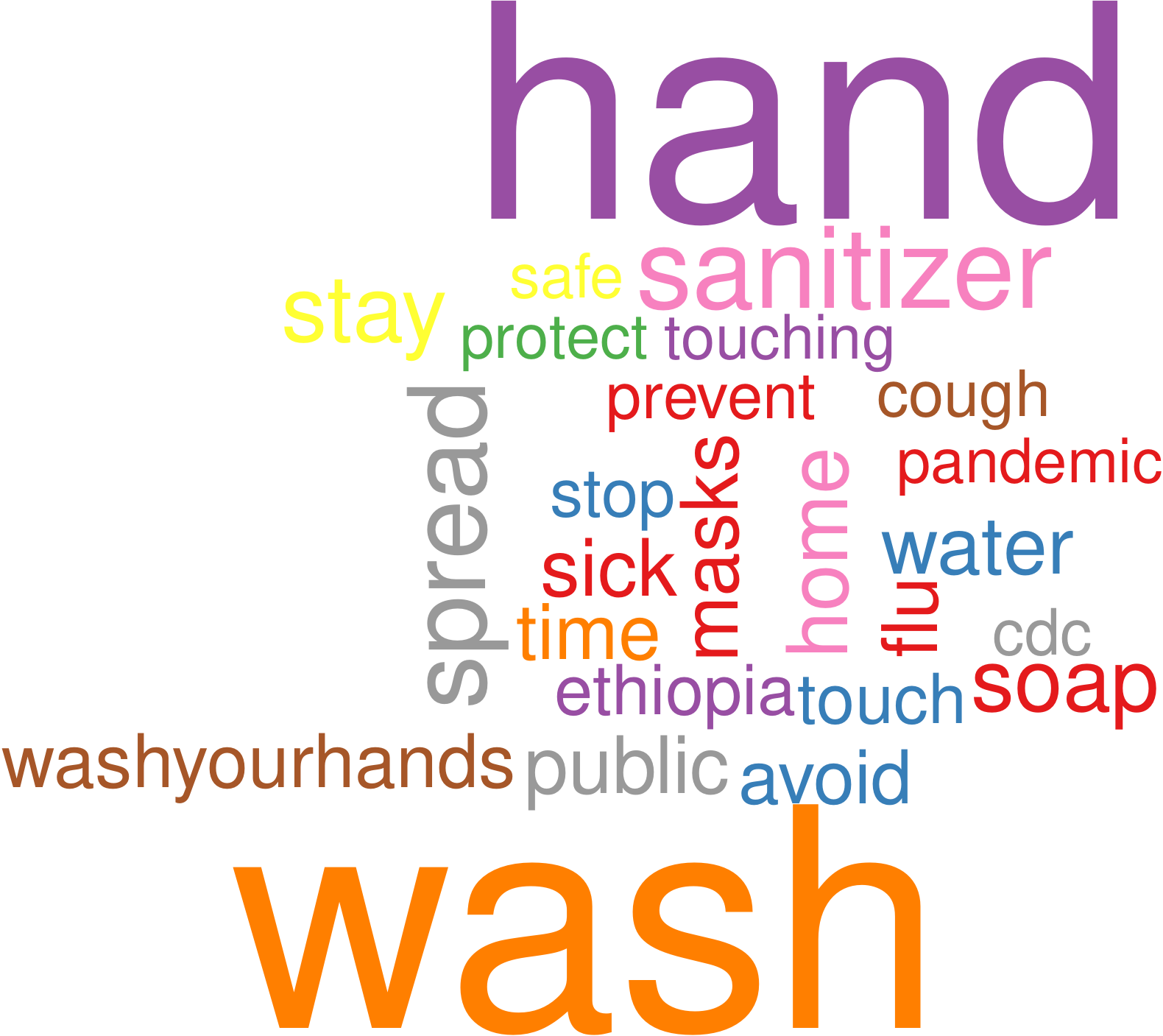}\label{fig:c}}
\hspace{3mm}
\subfloat[Travel]{\includegraphics[width=0.23\linewidth]{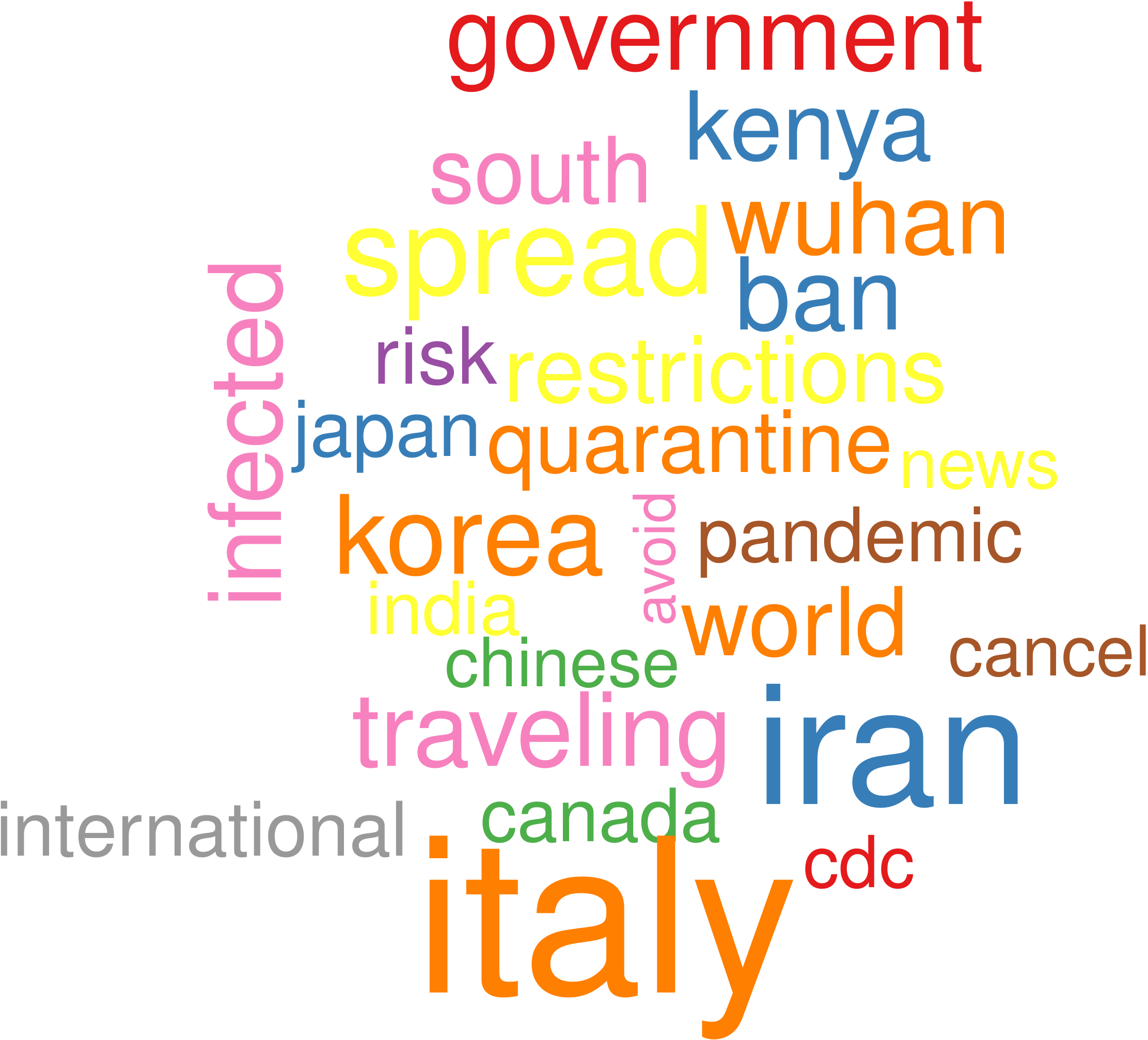}\label{fig:c}}
\caption{Wordcloud}
\label{Fig:Wordcloud}
\end{figure*}

\begin{figure}[ht!]
\includegraphics[width=\columnwidth]{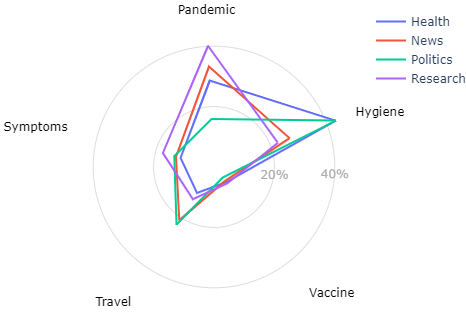}
\caption{Leaders alignment towards various public concerns.}
\label{Fig:Profession1}
\end{figure}

Tweets analysis belonging to the symptoms category shows that the daily percentage of observed COVID-19 tweets related to symptoms is higher than other concerns, reflecting that users on Twitter are more concerned about the disease symptoms. Furthermore, we extract the coronavirus symptoms from tweets (see Figure \ref{Fig:Wordcloud}a). The commonly discussed symptoms are fever, cough, and cold. Leaders also discuss the similarity of COVID-19 and influenza/flu, as both spreads among people in a similar way, that is, via respiratory droplets from coughing or sneezing \cite{rothan2020epidemiology}. Leaders call the COVID-19 an asymptomatic disease as, in many cases, an infected person does not show any symptoms \cite{rothan2020epidemiology}. As anticipated, the \textit{research} cluster is more concerned with understanding the COVID-19 symptoms than the other three clusters (see Figure \ref{Fig:Profession1}).

\subsection{Disease vaccination} 
Our dataset spans from February 01, 2020 to May 02, 2020 and in that time period no vaccination or specific treatment for COVID-19 was available, therefore, leaders are discussing the effectiveness of flu vaccination for COVID-19 as both flu and COVID-19 cause respiratory disease (see Figure \ref{Fig:Wordcloud}b). We can also observe that leaders are discussing intensely the ongoing research to cure COVID-19. Therefore, we can infer that leaders on Twitter are very conscious of the symptoms of COVID-19 and also keeping an eye on the vaccination research for COVID-19. As no vaccine was available in that time period, all clusters are cautiously tweeting regarding the vaccination (see Figure \ref{Fig:Profession1}).

\subsection{Disease countermeasures}
For more detail analysis, we partition countermeasures tweets into two categories: (i) \textbf{Hygiene} (using keywords such as \textit{hygiene, wash, hand}) and (ii) \textbf{Mask} (using keyword \textit{mask}). The explanation behind this categorization is that these two countermeasures (\textit{hygiene} and \textit{mask}) have distinct essence. We find that the volume of tweets pertaining to \textit{Mask} is higher than \textit{Hygiene} before February 24, and after that leader started focusing more on \textit{Hygiene} compared to \textit{Mask}. This indicates that initially, leaders were tweeting about wearing masks but later they shifted their countermeasures strategy towards taking proper \textit{Hygiene} against COVID-19.

To explore the countermeasures against COVID-19, we created a wordcloud from countermeasures category tweets (see Figure \ref{Fig:Wordcloud}c) indicating suggestions such as \textbf{handwashing}, \textbf{respiratory hygiene}, \textbf{self-isolation} and \textbf{self-quarantine} as prevention. Hand wash using soap and water or sanitizer is highly discussed for countermeasures. Handwashing is also recommended by the US Centers for Disease Control and Prevention (CDC) to prevent the spread of the disease. It recommends that people should wash hands more often with soap and water for at least 20 seconds, especially after going to the toilet or when hands are visibly dirty, before eating, and after blowing one's nose, coughing, or sneezing. It further recommended using an alcohol-based hand sanitizer with at least 60\% alcohol by volume when soap and water are not readily available\footnote{https://www.cdc.gov/coronavirus/2019-ncov/about/steps-when-sick.html}. The advice from CDC to avoid touching the eyes, nose, or mouth with unwashed hands and for \textit{respiratory hygiene}, cover the mouth with masks and take precautions in case of coughing\footnote{https://www.cdc.gov/coronavirus/2019-ncov/travelers/index.html} are also highly tweeted.

A high number of tweets belonging to \textit{Health} and \textit{politics} clusters are focused on hygiene compared to tweets from  \textit{news} and \textit{research} clusters (see Figure \ref{Fig:Profession1}).

\subsection{Disease transmission during travelling.}
To explore the effect of COVID-19 on traveling, we create a word cloud from travel category tweets (see Figure \ref{Fig:Wordcloud}d). This indicates that countries quarantined travelers from many countries to control pandemics or avoid spreading infection. Some of the mention countries either quarantined travelers from other countries or banned by others, such as China, Italy, South Korea, Japan, Iran, Canada, India, and Kenya. CDC has also issued guidelines for flight travelers and crew\footnote{https://www.cdc.gov/quarantine/air/managing-sick-travelers/ncov-airlines.html}. Among clusters, \textit{news} and \textit{politics} clusters are more focusing on travelling compared to \textit{research} and \textit{health} clusters (see Figure \ref{Fig:Profession1}). Although, it is interesting to observe that \textit{politics} cluster is discussing more about \textit{travel} compared to \textit{news}.

\subsection{COVID-19 as a pandemic}
As COVID-19 outbreak is first identified in Wuhan, China, in December 2019 \cite{world2020novel}. The World Health Organization (WHO) declared the outbreak a \textit{Public Health Emergency of International Concern} on January $30^{th}$ 2020 and a pandemic on March $11^{th}$ 2020~\cite{world2020novel2}. Our analysis on tweets shows that \textit{research, news} and \textit{health} clusters are frequently using pandemic or epidemic keyword while discussing about COVID-19 compared to \textit{politics} cluster (see Figure \ref{Fig:Profession1}). Interestingly, this reflects the non-supportive behavior of \textit{politics} cluster towards considering COVID-19 as pandemic \cite{rothwell2020politics}.

Next, we test the correlation of leading clusters with public concerns. Considering that both leading clusters and public concerns are categorical variables, we perform Chi square independent test. We begin with the null hypothesis ($H_0$) and alternate hypothesis ($H_a$) as:

\begin{itemize}
    \item $H_0$: The number of tweets related to public concerns is independent of the type of leading cluster.
    \item $H_a$: The number of tweets related to public concerns is dependent of the type of leading cluster.
\end{itemize}

The Chi square independent test results a p-value of 1.411e-33. Considering $\alpha=0.05$, we observe that p-value is smaller than $\alpha$. Since, $\alpha$ $>$ p-value, we reject $H_0$. This means that the factors are not independent. Hence, we can conclude that at a 5\% level of significance, from the data, there is sufficient evidence to conclude that the number of tweets related to public concerns and the type of leading cluster are dependent on one another.

To summarize, on analyzing the clusters' alignment towards various public concerns (see Figure \ref{Fig:Profession1}), we find that \textit{researchers} are highly concerned about understanding the COVID-19 by studying its symptoms and development of the vaccination. \textit{News} are discussing travel and hygiene. \textit{Health} cluster is focusing on hygiene. Whereas, \textit{political} people are highly concerned about travel and hygiene. This indicates that the different clusters are focused on specific public concerns. This can be viewed as a positive approach since various clusters are focusing on particular issues and engaging with each other on common problems.

\section{Tweet Classification In  Clusters}\label{Sec:Modeling}
Next, taking insights from previous sections, we build a model to estimate the likelihood that a tweet belongs to a specific cluster.

\subsection{Features used for learning}\label{subsec:features}
To illustrate the predictive power of various feature sets, we define a series of models, each with a different feature set, as mentioned in the earlier section. We focus on three different types of features. One of them is ``Tweet text'' is part of the original dataset features, and others (``Emotions'' and ``public concerns'') are extracted using tweet text. These features are the following:
\begin{enumerate}
    \item \textbf{Tweet text: }This refers to the clean text extracted after preprocessing the original tweets (see Section \ref{subsec:DataPreprocessing}).
    \item \textbf{Emotions: }This refers to the emotions associated with respect to each tweet (see Section \ref{Sec:Leaders}).
    \item \textbf{Public concerns: }This corresponds to the public concerns revealed in Section \ref{Sec:leadersConcern}.
\end{enumerate}

\begin{figure}
    \centering
    \includegraphics[width=\columnwidth]{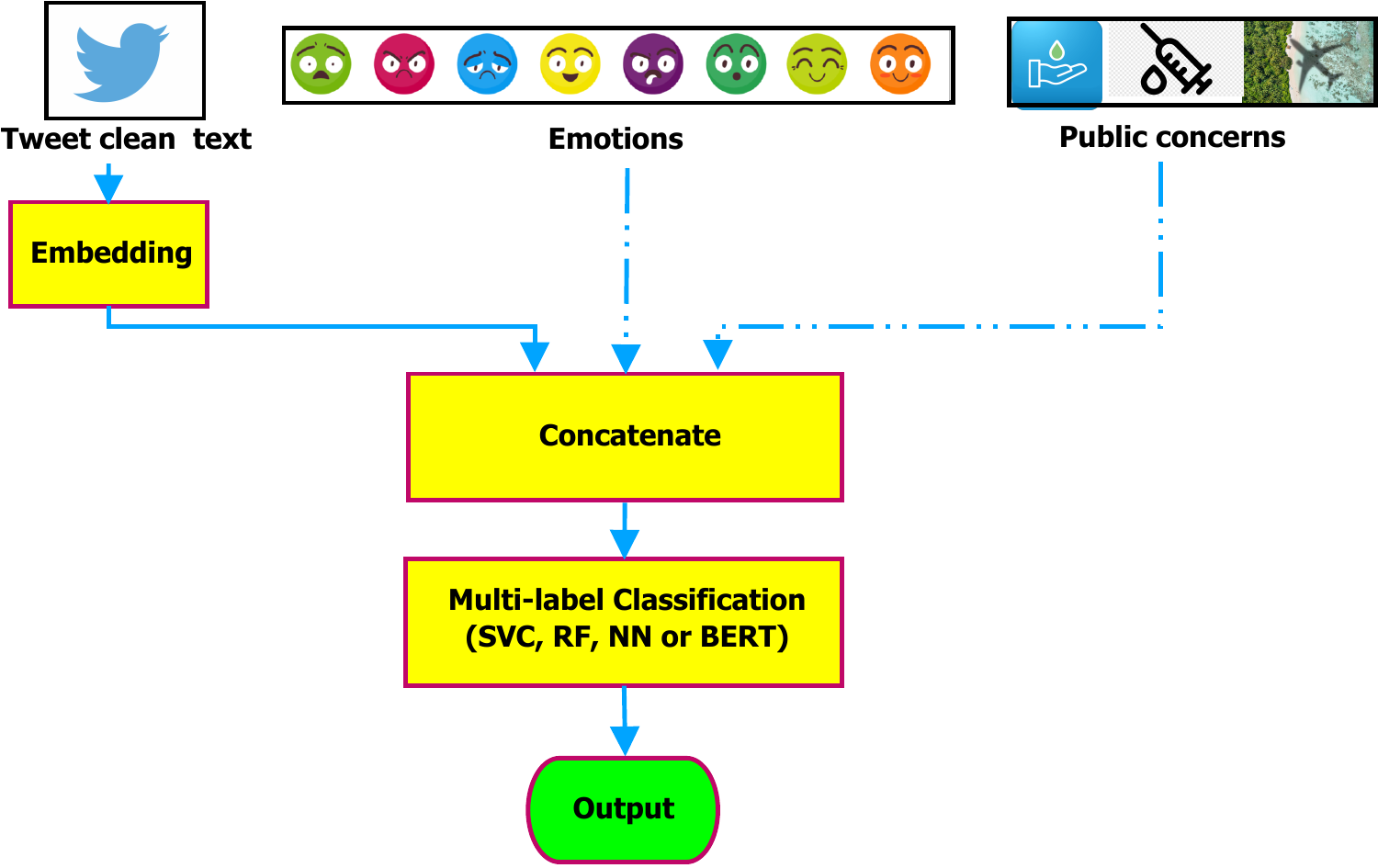}
    \caption{Flow diagram for features concatenation and model selection.}
    \label{fig:concate}
\end{figure}

\subsection{Experimental setup and results}
We aim to estimate the likelihood that a tweet belongs to a specific cluster using the features mentioned in Section \ref{subsec:features}. As in Section \ref{Sec:Leaders}, we filter the leaders and cluster them into four groups. We remove all user tweets that are not in any of the leader clusters. We also deleted all blank tweets after tweet preprocessing. After applying these filters, we obtain a dataset containing 42,468 tweets.

The tweets percentage and count belonging to different clusters are as follows: \textit{News} (36\%; 15,289), \textit{Health} (33\%; 14,023), \textit{Research} (18\%; 7,635) and \textit{Politics} (13\%; 5,521). As tweet distribution among clusters is imbalanced, this problem is represented as an \textit{imbalanced multi-class classification} \cite{zhang2015towards}. Since the dataset is imbalanced, we use Synthetic Minority Over-sampling Technique (SMOTE) \cite{chawla2002smote} to resolve this problem. SMOTE works by selecting examples close in the feature space, drawing a line between the examples in the feature space, and drawing a new sample at a point along that line. Specifically, a random example from the minority class is first chosen, then k of the nearest neighbors for that example is found (typically k=5). A randomly selected neighbor is chosen, and a synthetic example is created at a randomly selected point between the two examples in feature space.

\begin{figure}
    \centering
    \includegraphics[width=\columnwidth]{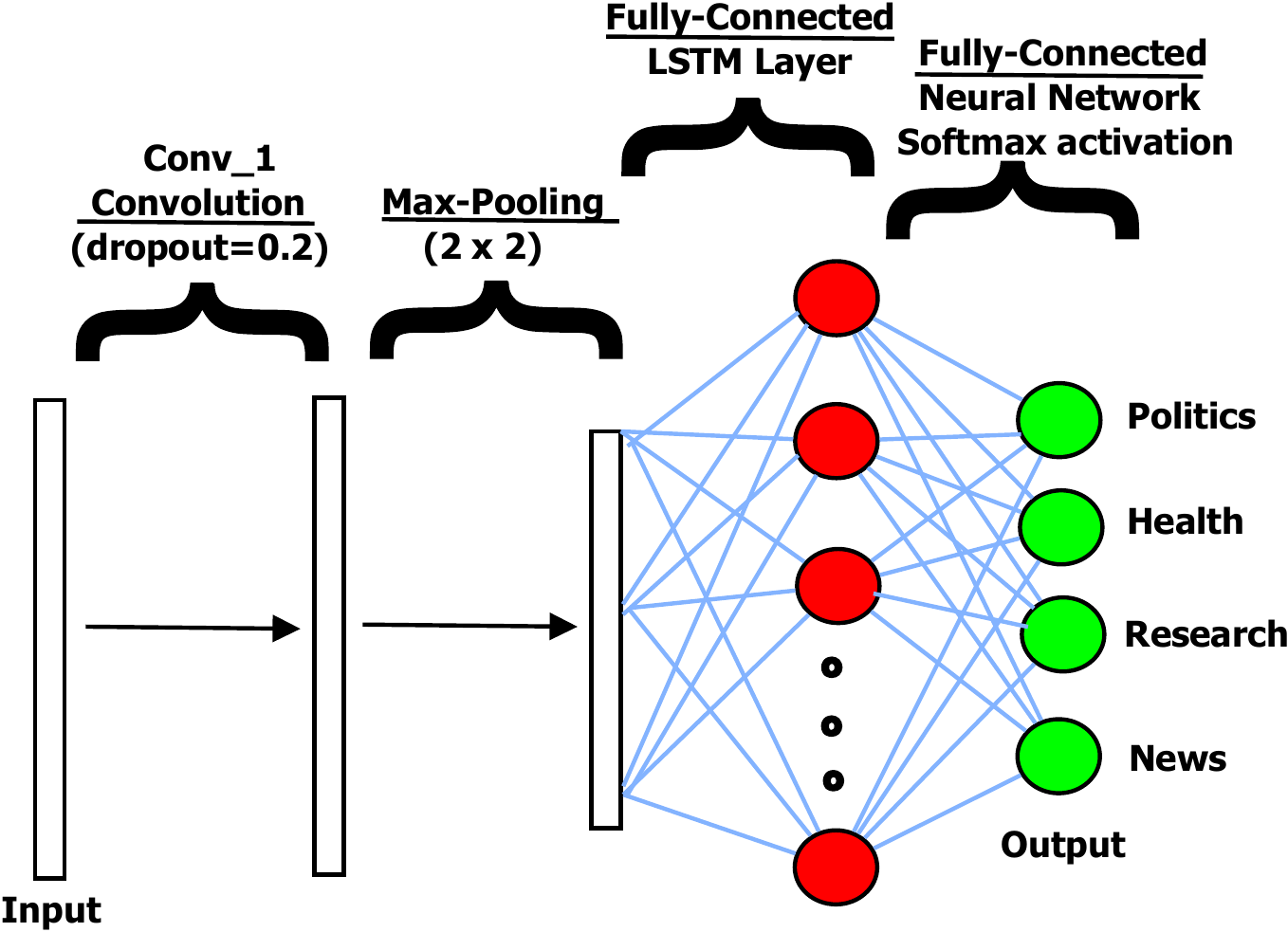}
    \caption{Random neural network framework.}
    \label{fig:NN}
\end{figure}

We experiment with several classification models, including Support Vector Classifier (SVC) \cite{tsuda1999support,lee2007domain}, Random Forests (RF) \cite{breiman2001random}, Random neural network (NN, see Figure \ref{fig:NN} for framework) \cite{hansen1990neural,alom2020deep}, and Bidirectional Encoder Representations from Transformers (BERT) \cite{devlin2018bert}. Figure \ref{fig:concate} displays the data flow to the models. We find that the Random Forest model is the most effective for our task. As the dataset is imbalanced and the trade-off between true and false positive rates associated with classification, we choose to compare models using the area under the receiver operating characteristic (ROC) curve (AUC) \cite{davis2006relationship,flach2004many}. Thus, a random baseline will score 50\% on AUC ROC. We use 5-fold cross-validation for estimation and we run all models three times. Therefore, 15 iterations in total for each model. We also standardize \textit{emotion} and \textit{public concern} features to have zero mean and unit variance.

\textbf{Results: }Table \ref{tab:result} shows the classification accuracy for our models. With the RF model trained on all available features, we achieve a mean accuracy of 96\% AUC ROC with 0.2\% as standard deviation. The low value of standard deviation indicates that the model is robust. All model shows that the most crucial feature for the classification is clean text. We can note that our best model, RF, is trained with the TF-IDF embedding for the words a user used in their tweets. We also see that public concerns and emotions are essential characteristics for classifying a tweet in different clusters.

\begin{table}
\begin{center}
\begin{tabular}{|p{0.12\textwidth}||*{4}{c|}}\hline
Data
&\makebox[3em]{SVC}&\makebox[3em]{RF}&\makebox[3em]{NN}
&\makebox[3em]{BERT}\\\hline\hline
Text & 91.9(2) & 95.3(2) & 91.8(9) & 92.5(8) \\\hline
Text+Concerns & 92.4(3) & 95.3(2) & 92.7(9) & 92.8(9) \\\hline
Text+Emotions & 93.2(3) & \textbf{96.0(2)} & 93.5(9)  & 93.6(8) \\\hline
Text+Emotions + Concerns & \textbf{93.6(2)} & \textbf{96.0(2)} & \textbf{94.1(9)} & \textbf{95.2(8)} \\\hline
\end{tabular}
\caption{Model mean AUC ROC and standard deviation for various dataset features. For example, the BERT model with the dataset (Text+Emotion) achieves 93.6\% mean AUC ROC accuracy with 0.8\% standard deviation.}
    \label{tab:result}
\end{center}
\end{table}

\section{Discussion And Conclusion}\label{Sec:Conclusion}
With a quest to understand the role of various leaders during COVID-19, we study a large number of tweets using techniques such as network analysis, text analysis, and sentiment analysis. Based on network analysis, users are categorized into four different clusters: \textit{research}, \textit{news}, \textit{health}, and \textit{politics}. The results using text analysis shows that leaders of different clusters are focused on various public concerns. In particular, researchers about understanding the pandemic \textit{symptoms} and development of \textit{vaccination}; news about \textit{travel} and \textit{hygiene}; health cluster about \textit{hygiene}; and political individuals about \textit{travel} and \textit{hygiene}. Additionally, sentiment analysis indicated that emotions such as \textit{anticipation}, \textit{fear}, \textit{sadness}, and \textit{trust} dominate all clusters. Also, the evolution of emotions with time shows  that there is only minor changes in emotions of all clusters, except in the politics cluster. In February, the politics cluster were showing more \textit{Trust}, but over time, it changed into \textit{Sadness} and \textit{Anticipation}. Lastly, we showed that the extracted features could be used to identify tweets' clusters with an AUC ROC score of up to 96\%.

\textbf{Limitations.} This work has some limitations, primarily related to generalization and users' response analysis on tweets. First, the analysis and classification model may not be generalized for the data collected in similar epidemic/pandemic situations (such as \emph{\#Ebola}). Our analysis is based upon users' posts on Twitter (i.e., tweets). Ideally, it would help us analyze the comments (that is, reply and retweets) regarding original tweets to get a broader understanding of the topic. Comments can be helpful to understand users' reactions to tweets. However, we showed that even an existing LDA model could effectively extract different topics from tweet text. Another limitation of our work is that the Twitter stream is filtered following Twitter's API documentation; hence the tweets analyzed here still constitute a representative subset of the stream instead of the entire stream. 

\textbf{Future work} could consider several important directions to determine the effect of users' response and media (such as images and videos) in such pandemic situations. We intend to analyze the tweets' data to a greater extent for different category users over time to understand their change in tweet pattern and focused concerns. Future work should also explore more sophisticated models and deeply analyze various cluster writing styles to capture their tweets' signature.

\section{Declarations}
\subsection{Funding}
This research is funded by ERDF via the IT Academy Research Programme, and H2020 framework project, SoBigData++, and CHIST-ERA project SAI.
\subsection{Conflicts of interest/Competing interests}
The authors declare that they have no conflict of interest.
\subsection{Availability of data and material}
Our dataset is Twitter data, and it is not allowed to share this dataset publicly according to Twitter data sharing rules.
\subsection{Code availability}
Not applicable. As  this  work  utilizes  a  non-shareable dataset.
\subsection{Authors' contributions}
\textit{Rahul Goel:} Data curation, Formal analysis, Investigation, Validation, Visualization, and Writing – original draft.

\noindent \textit{Rajesh Sharma:} Supervision, and Writing – Review \& Editing.

\bibliographystyle{spmpsci}
\bibliography{main}

\end{document}